\documentclass[12pt]{article}

\usepackage{epsfig}
\usepackage{cite}
\usepackage{amsmath, amssymb, amsfonts}
\usepackage{color}
\usepackage{latexsym}
\usepackage{graphicx}
\usepackage{cancel}
\usepackage[colorlinks,bookmarks]{hyperref}
\hypersetup{pdfpagemode=UseNone, pdfstartview=FitH, linkcolor=blue,
            citecolor=red, urlcolor=blue}

\def\be{\begin{equation}}
\def\ee{\end{equation}}
\def\ba{\begin{eqnarray}}
\def\ea{\end{eqnarray}}

\def\bdm{\begin{displaymath}}
\def\edm{\end{displaymath}}

\def\bq{\begin{quote}}
\def\eq{\end{quote}}

 at 10truept

\newcommand{\bea}{\begin{eqnarray}}
\newcommand{\eea}{\end{eqnarray}}

\newcommand{\bi}{\begin{itemize}}
\newcommand{\ei}{\end{itemize}}

\newcommand{\beq}{\begin{equation}}
\newcommand{\eeq}{\end{equation}}
\newcommand{\beqa}{\begin{eqnarray}}
\newcommand{\eeqa}{\end{eqnarray}}

\def\12{{1 \over 2}}

\def\ltap{\ \raise.3ex\hbox{$<$\kern-.75em\lower1ex\hbox{$\sim$}}\ }
\def\gtap{\ \raise.3ex\hbox{$>$\kern-.75em\lower1ex\hbox{$\sim$}}\ }
\def\gl{\ \raise.5ex\hbox{$>$}\kern-.8em\lower.5ex\hbox{$<$}\ }
\def\roughly#1{\raise.3ex\hbox{$#1$\kern-.75em\lower1ex\hbox{$\sim$}}}

\begin{document}

\thispagestyle{empty}
\begin{flushright}
May 2023 
\end{flushright}
\vspace*{1.5cm}
\begin{center}
  
{\Large \bf Origin of the Arrow of Time in Quantum Mechanics}

\vspace*{1.55cm} {\large Nemanja Kaloper$^{b, }$\footnote{\tt
kaloper@physics.ucdavis.edu}
}\\
\vspace{.5cm}
{\em $^b$QMAP, Department of Physics and Astronomy, University of
California}\\
\vspace{.05cm}
{\em Davis, CA 95616, USA}\\

\vspace{1.5cm} ABSTRACT
\end{center}
We point out that time's arrow is naturally induced by quantum mechanical evolution, whenever the systems
have a very large number ${\cal N}$ of non-degenerate states and a Hamiltonian bounded from below.
When ${\cal N}$ is finite, the arrow is imperfect, since evolution can resurrect past states.
In the limit ${\cal N} \rightarrow \infty$ the arrow is fixed by the ``tooth of time": the decay of excited states induced by {\it spontaneous emission} 
to the ground state, mediated by interactions and a large number of decay products which carry energy and information to infinity. This applies to 
individual isolated atoms, and does not require a coupling to a separate large heath bath. 

\vfill \setcounter{page}{0} \setcounter{footnote}{0}

\vspace{1cm}
\newpage

\section{Introduction}

Time seems to have an arrow -- a `flow' direction, which gives a meaning to the causal order of events. 
Yet in microphysics, many local microphysical laws are time-reversal invariant,
implying that microscopic arrow of time is absent. This is often taken to imply that time's arrow originates macroscopically, and perhaps from cosmology. An example
would be the Second Law of Thermodynamics. The precise reason behind the Second Law is Boltzmann's kinetic theory and interactions between many 
microscopic constituents. The Second Law states that physical systems typically evolve toward most common states, 
and this evolution can be taken to define time's arrow.
Although this makes sense, it is a leap of faith to declare that the 
fundamental time's arrow emerges together with the universe because of entropy considerations. If so, this could force a
very lonely observer in some very cold and dark part of the universe to `know' of the time's arrow {\it despite} the lack of sharing experiences with the rest of the world. Given the
notion of decoupling -- loosely, that phenomena correlate only if they can interact with each other -- this should seem disturbing to us. Hence here we will attempt to understand this phenomenon from a more local point of view.  
For a general sampling of ideas, suggestions, confusions and criticisms concerning time and its arrow, see \cite{Gold:1958koa,Penrose:1979azm,Davies:1983nf,Page:1983uh,Davies:1984qc,Penrose:1988mg,Albrecht:2002uz,Carroll:2004pn}
and references therein. Some complementary ideas to those expressed here, we think, 
are shown in e.g. \cite{Yukalov,Donoghue:2020mdd,Frohlich:2022yir}. 

In what follows we will try to understand how time enters in quantum mechanics, for the most part ignoring gravity and cosmology. 
The so called ``problem of time" \cite{Isham:1992ms}, although recently mostly invoked in the context of (quantum) cosmology, has been contemplated in 
quantum mechanics, and so we believe this is a good starting point. In quantum mechanics we normally 
think of time as it appears in Schr\"odinger equation, parameterizing the infinitesimal translations generated by the Hamiltonian \cite{messiah}. 
This ``background" time, extrinsic to the system, is introduced by an external observer as a reference. That observer must have an access to a different system -- a classical clock -- which defines $t$ and its flow. Yet this time is introduced 
entirely by hand, by writing the Schr\"odinger equation in its usual form without specifying where the background time $t$ 
comes from, or what it represents. 
It is natural to do so when transitioning from classical to quantum wave mechanics. Time is ``inherited" from the classical realm, and ushered in 
by correspondence principle. 

Here we will argue that there is a sense where, from a point of
view intrinsic to the quantum system being described, this time can be thought of as being `emergent': the fact that it appears in Schr\"odinger equation 
does not mean it's ``there" - i.e. that it's really physically relevant per se. The same point generally arises in the presence of continuous symmetries. A  
variable which depends on a parameter that runs along an orbit of a symmetry group does not actually change by varying the orbit parameter. The variable 
naturally resides in the coset space of the symmetry group, and a physical variation corresponds to the transition from one orbit to another. 
In our case, in simple terms, for the background time to be physically real, 
the system must be able to ``see" its presence. In other words, there must be physical observables intrinsic to the
system which evolve in correlation with $t$ and its arrow. Otherwise, if no observables change in a way that is correlated with the variation of $t$
the background time is  good-for-naught. We suggest that the situation is really 
equivalent to what one encounters in canonical quantum gravity, where the ADM Hamiltonian density is identically zero 
by general covariance \cite{Arnowitt:1962hi}. In this case, the functional Schr\"odinger equation reduces to
${\cal H}_{ADM} \Psi = 0$, and one does not even have a clear path of introducing a background time, let alone its arrow. 
Nonetheless, the main issue may not be that the Hamiltonian vanishes, but that the Hilbert space is totally degenerate. 

It is known that there is no intrinsic time variable in standard quantum mechanics, as there is no time-operator, whose eigenvalue 
time can be, and which is intrinsically `induced' by system's own dynamics. This is a corollary of the work of 
\cite{Susskind:1964zz,Carruthers:1968my}, who prove that for 
a system on a circle there does not exist an `angle' operator which has the angle variable as an eigenvalue. The periodicity of the angular variable $\phi$ implies that the angular momentum operator $L_\phi$, conjugate to $\phi$, has a spectrum bounded from 
below, with self-adjoint boundary conditions. Since the Hamiltonian and the background time $t$ are in the same correspondence 
as $L_\phi$ and $\phi$, the same conclusion follows for $t$. While there is no doubt that the parameter $t$ in Schr\"odinger 
equation plays the role of time well in many cases, it and its orientation are extraneous to the system dynamics,
and when they appear in the description of the dynamics they can be viewed as being ``emergent". 
We will try to shed some light on why this is so. We will also argue that once time has emerged, its orientation -- i.e. ``time's arrow" -- will emerge whenever 
the quantum system has a unique ground state. We will use as the example the Hydrogen atom, but the conclusions are more general. The main
point here, which we are trying to make, is that the problem of the origin of time's arrow seems -- to us -- to be secondary 
to the ``problem of time" -- and that
it may be far easier to solve the former once the latter is sorted 
out. This is in stark contrast to, for example, views expressed in \cite{DiBiagio:2020jbd,Rovelli:2021elq}, where it is asserted that all irreversible 
phenomena in nature are macroscopic and statistical. Instead, we argue here that even a simple microphysical system such as an
isolated Hydrogen atom knows of time's arrow. While our argument is entirely based on known calculations we are unaware of the conclusion that
we offer, which is that microscopic evolution of isolated quantum systems with few ``moving parts" picks a direction of time, except for the paper
\cite{Buchholz:2023lhl} which appeared within days of this work, and which argues for the same conclusion.  
At the very end we will also comment on some possible implications
for cosmology, albeit without much rigor and detail. 

\section{Time's Arrow and Quantum Mechanics}

We start with a quantum-mechanical system with all states which are completely degenerate in energy. We will take the number of states to be finite without loss of generality\footnote{We may realize such a
system as a subsystem of a larger system with a large symmetry group, by restricting to only the states which comprise a complete irreducible orbit 
of the symmetry group.}. Since the time evolution is controlled by the Schr\"odinger equation ($\hbar = 1$),
\be
i \partial_t  | \psi \rangle = H | \psi  \rangle \, ,
\label{schro}
\ee
and the states are completely degenerate,
\be
H  | \, {\rm any} \, \rangle = E  | \, {\rm any} \, \rangle \, ,
\label{degham}
\ee
there is no real notion of time in the system, despite the presence of the backrgound time parameter $t$ in the Schr\"odinger equation. 
Indeed, take any linear combination of states,
\be
| \psi  \rangle = \sum_{n} c_n | \psi_n \rangle \, , 
\ee
and the ``background" time evolution is always the same for any such state,
\be
| \psi (t)  \rangle = e^{-i E(t-t_0)} | \psi (t_0) \rangle \, , 
\ee
no matter what. 

Now, the background time $t$ is introduced here `technically' by an observer with the knowledge of the eigenvalues of the Hamiltonian. 
In terms of this background time, in this case we can say that system is time independent -- it does not change as the external, 
background time progresses. The overall phase factor is irrelevant because of the phase shift symmetry, as long as the 
Hamiltonian does not change with background time.  Because we can only measure phase differences in quantum mechanics, and not
absolute phases, in degenerate systems the phase evolves in exactly the same way for all states and so the phase difference is zero,
and time dependence drops out. 

This allows us to reinterpret `time independence' as the statement 
that in a toy model universe which is only comprised of this system, time independence is intrinsically 
indistinguishable from the statement that time does not exist! Note, that this is completely equivalent to what one often encounters in canonical quantum gravity, where by general covariance the ADM Hamiltonian density is identically zero \cite{Arnowitt:1962hi}. 
In that case, the functional Schr\"odinger equation reduces to
${\cal H}_{ADM} \Psi = 0$, which has identical practical consequence for the nature of time, as does degeneracy of energy levels in our 
toy quantum system. After all, without gravity, we are free to shift the Hamiltonian by a constant, and so we can define 
$H' = H-E$ which vanishes on all states in our degenerate system. 
So from the point of view intrinsic to the system, using this map of the degenerate quantum system to the familiar situation in 
quantum cosmology, we can conclude  
that time does not really exist in systems with completely degenerate Hilbert spaces,  
because quantum mechanics is a theory of ray representations and the total overall phase is not an observable. 
We can introduce the background time, but it is of no consequence: there is no physical distinction between eternal 
constants and absence of time, be they zero or not. 

In particular, the probability to find the system in any particular state 
$|\hat \psi \rangle$, and at any particular time $t$, is independent of $t$ since 
\be
|\langle \hat \psi | \psi(t) \rangle|^2 = |\langle \hat \psi | \psi(t_0) \rangle|^2  \, .
\ee
The same is true for any observable in completely degenerate systems, where observables are constructed from ingredients intrinsic
to the system. There is no operational notion 
of time, nor of its arrow: $t$ cannot be measured 
(by noting any particular order of events), and so it may as well not exist in physical reality of this system. In a manner 
of speaking, the system completely ignores the background time $t$ -- which is a good
sign it might as well not exist\footnote{In this case, the time $t$ needs the external observer, 
and whatever other observation this one can carry out, in order to be reaffirmed. Short of that, we might as well drop it.}. 

If we relax the condition of looking at a system with exactly degenerate states, time shows up. 
Again, we stress that while the background time $t$ always `exists', as introduced by some 
external observer, it is completely extraneous unless the system has a means of `measuring' it. 
For example, let us extend the degenerate 
system above with the inclusion of a single additional state $| \phi \rangle$, with energy ${\cal E} = E + \epsilon$. In this case
if we take a general state  
\be
| \Psi  \rangle = a | \phi \rangle + \sum_{n} a_n | \psi_n \rangle = a | \phi \rangle + | \psi \rangle  \, , 
\ee
time evolution is non-trivial when $a \ne 0$:
\be
| \Psi  (t) \rangle =  e^{-i {\cal E}(t-t_0)} a | \phi (t_0) \rangle + e^{-i E(t-t_0)}  | \psi (t_0) \rangle \, .
\label{evo}
\ee 
The observables now depend on $t - t_0$, as exemplified by probabilities. Although the projections of
$|\Psi(t) \rangle$ onto the mutually orthogonal subspaces $|\psi \rangle$ and $|\phi \rangle$ are still constant, as are
any other observables strictly localized to a subspace, there are cross terms. The state $| \Psi  (t) \rangle$ spins through
the Hilbert space and its projection onto some arbitrary state  $|\hat \Psi \rangle$ which has support in both subspaces $|\psi \rangle$ and $|\phi \rangle$ 
is, using Eq. (\ref{evo}), 
\be
\langle \hat \Psi | \Psi(t) \rangle = b^* \langle \phi  | \Psi(t) \rangle + \langle \hat \psi | \Psi(t) \rangle 
=  ab^* e^{-i {\cal E}(t-t_0)} +  e^{-i E(t-t_0)}  \langle \hat \psi | \psi \rangle  \, .
\ee
Thus the probability is
\be
|\langle \hat \Psi | \Psi(t) \rangle |^2 
=  |ab^*|^2 + |\langle \hat \psi | \psi \rangle|^2 +   ab^* \langle \hat \psi | \psi \rangle^* e^{-i [{\cal E}-E](t-t_0)} 
+ a^* b \langle \hat \psi | \psi \rangle e^{i[{\cal E}-E](t-t_0)}  \langle \hat \psi | \psi \rangle  \, .
\ee
Introducing the moduli $\rho = |ab^*|$ and $\sigma = |\langle \hat \psi | \psi \rangle|$ and the relative phase between the constant coefficients 
$\delta = {\tt Arg}(ab^*) - {\tt Arg}(\langle \hat \psi | \psi \rangle)$, and using  ${\cal E} = E + \epsilon$, we can rewrite this equation as
\be
|\langle \hat \Psi | \Psi(t) \rangle |^2 
= \rho^2 + \sigma^2 +   2 \rho \sigma \cos[\epsilon(t-t_0) + \delta]  \, .
\ee
The (non-normalized) probability now oscillates between $(\rho + \sigma)^2 $ and $(\rho - \sigma)^2$, with the period $T = 2\pi/\epsilon$.
When $\epsilon \ll 1$, the period is very long, and the probability lingers around the extrema for a long time. 

The `probability waves' above behave just like ``beats" in classical mechanics: the superposition of two waves with slightly different frequencies, which interfere together. As the system evolves, the two waves come in and out of being in phase, and develop an evolving pattern of maxima and minima\footnote{A good visual example is the oscillations of an image of a pendulum which oscillates in front of an oscillating mirror.}. Near the extrema of their superposition, 
the system remains in or out of phase for a long time when $\epsilon$ is small, and constructive or destructive interference 
can be very efficient\footnote{In fact, this is a ``primitive" variant of the phenomena of quantum collapse and revivial, which for example 
appear in the finite-dimensional
Jaynes-Cummings model \cite{Jaynes:1963zz} of quantum optics, as noted in \cite{Eberly:1980zz}. In cosmological conditions analogous physics can occur in de Sitter \cite{Dyson:2002pf}.}. To make the analogy with ``beats" exact we can introduce a third subsystem, nearly degenerate with the other two. This is
because the overall phase cancels in a ray representation. In particular, if the individual contributions from different energy eigenfunctions are
equally weighted, 
the system looks like a state which is periodically appearing or disappearing from
reality. 
\begin{figure}[thb]
    \centering
    \includegraphics[width=1\textwidth]{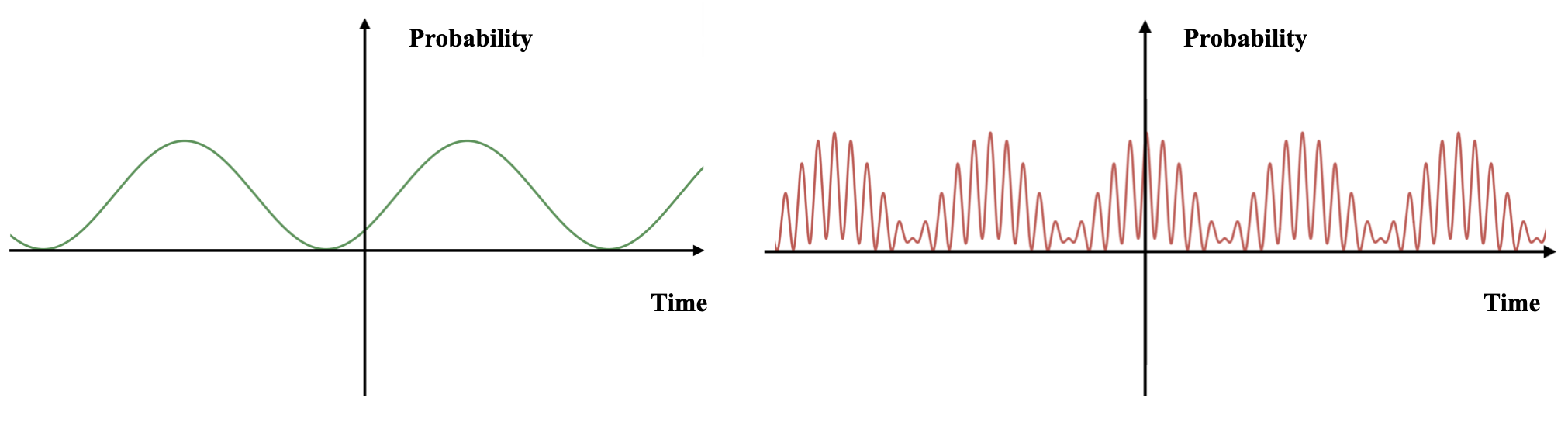}
          \caption{Quantum clocks with equally weighted eigenfunctions and two (left panel) or three (right panel) energy levels.}
    \label{fig1}
\end{figure}

A process like this can be used as a clock. Time has now emerged\footnote{As we stressed above, time is not an eigenvalue of an operator as long as the Hamiltonian is bounded from
below, as explained in \cite{Susskind:1964zz}. This does not preclude emergent time, but it explains why quantum systems do not have an intrinsic time variable,
arising as an eigenvalue of an observable.}: the system appears to evolve relative to the background time $t$, such that observables determined by an external observer, or by some constituent of the system itself, do seem to be different and different moments of $t$. Nevertheless, the system still does 
not have a definite arrow. We can pick the direction of the flow of the background time $t$ at will, making a single discrete choice. But, 
we can reverse it at will, reversing also the order
of events, picking the arrow either way. Moreover, the evolution will do it itself, by recurrences in the finite phase space. 
E.g. if the oscillations are slow, we can pick an arrow of time
and follow the evolution of the system for a long time, but eventually we will encounter the return to the initial configuration, and decide that the time's arrow
is not ``conserved". An arbitrarily chosen state of the system, having contributions from only a 
finite number of energy levels, will in principle regularly scan between its constituents, analogously to the two-level example we have given here. Arrow of time will seem to flip-flop. If more
states and energy levels are added to the mix, the time intervals for both the quantum collapse (reduction of the oscillation amplitude) and the quantum revival (the regrowth 
of the oscillation amplitude) can be controlled, squeezing and stretching the ``wave packets". But the general conclusions remain. 

Let us now consider the case when the system has infinitely many different energy 
levels\footnote{In fact, a full continuum of them -- the full blown quantum electrodynamics.}. If these are all exact eigenstates of the Hamiltonian,
we can set up states which will cycle between different constituents regularly, as if the number of levels were finite. We are not interested in such cases,
since they do not really add anything new to the discussion. Instead, to define what we are interested in, let us consider the example provided by Hydrogen atom, and its 
dynamics. In non-relativistic quantum mechanics, Hydrogen atom is described by the Hamiltonian
\be
H = - \frac{1}{2m} \vec \nabla^2 - \frac{\alpha}{|\vec r \, |} \, , 
\label{hamhyd}
\ee
and has infinitely many discrete bound states, whose energies are 
\be
E_n = - \frac{R}{n^2} \, , ~~~~~~ n = 1, 2, 3, \ldots
\ee
where $R = m \alpha^2/2$ is the Rydberg constant
for Hydrogen atom, and $\alpha = e^2/4\pi \simeq 1/137$ the fine structure constant. 
Each level is $2n^2$ times degenerate, due to spin and the conserved Lenz vector, which enhances the manifest $O(3)$ symmetry to $O(4)$. 
At first glance, since $H$ in (\ref{hamhyd}) is Hermitean, the energies $E_n$ are real, and so the eigenstates $| \psi_n \rangle$ are absolutely stable. If we took any individual energy eigenstate,
it would never change by Hamiltonian-induced translation. Likewise, if we took a generic 
linear combination of them as an initial state, it would seem to cycle between different constituents without an end -- the situation 
which we just declared a lack of interest in. 

\underline{However, this is incorrect.} The standard textbook description of Hydrogen in nonrelativistic limit is just a nonrelativistc {\it approximation}. 
The point is, that the full quantum state of Hydrogen must also specify the state of photons, not 
only the electron\footnote{The proton, being heavy, beside making $m$ in (\ref{hamhyd}) 
the reduced mass, is just an inert spectator.}, since 
its charge is nonzero and the photon is massless. A generic state of the atom is given by a linear combination of direct product states 
$\propto | {\rm electron} \rangle \otimes | {\rm photon} \rangle$. Importantly, the photon states 
are {\it not} the photon vacuum. 

We can describe what happens by recalling that Coulomb potential in the Hydrogen Lagrangian comes from the timelike components 
in the covariant interaction
\be
A_\mu \bar \psi \gamma^\mu \psi = A_0 \psi^\dagger \psi + \ldots \, , 
\ee
in nonrelativistic limit. The vector $A_\mu$ is the quantized photon field, and the Coulomb potential is 
its expectation value between in- and out-photon states. If in perturbation theory the photon state were an empty vacuum 
of $A_\mu$,  $\langle 0 | A_\mu | 0 \rangle$ would be zero. Instead, the photon state in Hydrogen atom must be a coherent state 
$| {\cal C} \rangle \propto \sum_n c_n | n~{\rm photons} \rangle$ \cite{Glauber:1963tx,Klauder:1995yr}, such that 
$\langle {\cal C} | A_0 | {\cal C} \rangle = - {\alpha}/{|\vec r \, |}$. Thus the ground state of Hydrogen involves all the
virtual photons whose squeezing into coherent states to leading order results in the Coulomb potential. 
Once this is accounted for, the full gauge invariant 
perturbation theory description of Hydrogen requires replacing $-i \vec \nabla \rightarrow - i \vec \nabla + e \vec A$, and the 
vector potential fluctuation can be taken to be in the Coulomb gauge, $\phi = 0$, $\vec \nabla \cdot \vec A = 0$. The operator
$\propto \vec A \cdot \vec \nabla$ mediates the decay of excited states to the ground state, triggering spontaneous emission. 

This phenomenon is purely quantum-mechanical. A variant of it has been found 
experimentally by L. Meitner \cite{meitner} (in nuclear physics) and, independently, P. Auger \cite{auger} (in atomic physics). 
The precise theory of it is based on the article 
by Dirac which initiated the foundation of quantum field theory \cite{Dirac:1927dy}, which showed the decay rate, as confirmed shortly later
in the papers \cite{Weisskopf:1930au,Weisskopf:1930ps} by Weisskopf and Wigner. A nice summary can be found in \cite{lamb,gerryknight}.

And yet the main result was correctly calculated by Albert Einstein, almost a decade prior, before the glimpses of quantum field theory, 
and before the experimental discoveries of Meitner and Auger. Einstein presciently \underline{\it anticipated} the phenomenon of spontaneous emission \cite{Einstein:1917zz} in 1917, in his rederivation of 
Planck's black body spectrum of radiation. We summarize this remarkable work here, following delightful lectures by Tong \cite{tong}. 

Einstein started with a system of atoms immersed in a reservoir of photons, and assumed that the system is in thermal equilibrium. 
He allowed that atoms and photons interact, and that atoms can both absorb and emit photons, by having an electron change 
the level. He further observed that the absorption and emission are amplified in environments where the radiator is surrounded 
by photons, but also included a channel where an excited, higher energy, atom state can decay into a lower energy, ground, state,
by releasing a photon even in the empty reservoir. In ter Haar's translation of Einstein's original article \cite{Einstein:1917zz},
Einstein stated

\vskip.3cm
\noindent {\it Accordingly, let it be possible for a molecule to make {\bf without external stimulation}\footnote{Boldface by the author.} 
a transition from the state $Z_m$ to the state $Z_n$ while emitting the radiation of energy $\varepsilon_m - \varepsilon_n$ of frequency
$\nu$.}
\vskip.3cm

\noindent Two lines below, Einstein introduced his famous $A$-coefficient, measuring the rate of this transition. The inclusion 
of this parameter opened up the possibility of spontaneous emission, with the calculation to provide the check if it is nonzero. The reverse process, of spontaneous excitation,
was ignored from the outset, as it were. Ignoring it is fully justified: if there were no photons anywhere near an atom in the
ground state, it could self-excite only by emitting a photon of $ \omega < 0$ -- an obvious affront to physics to and fro (and a reminder
that Hydrogen Hamiltonian is bounded from below). 

As noted above, Einstein also added the stimulated emission and absorption processes, whereby the atoms emit and absorb into/from the 
environment photon population, which are proportional to the number of environment photons per unit volume $n_\gamma$, 
and occur between excited and ground states with energies
$E_e - E_g = \omega$. The subscripts $e$ and $g$ refer to the excited and ground state. As noted above, we take all units to be
natural, so $\hbar = k_{\tt B} = c = 1$. Using the principle
of detailed balance for reactions, with the probability for the emission 
and absorption of a single photon being $B_{eg}$ for an excitation (the transition $g \rightarrow e$) and $B_{ge}$ for a relaxation ($e \rightarrow g$),  
\be
\delta N_{e} = n_\gamma(\omega) \Bigl(B_{eg} N_g - B_{ge} N_e \Bigr) - A_{ge} N_e \, , ~~~~~~
\delta N_{g} = - n_\gamma(\omega) \Bigl(B_{eg} N_g - B_{ge} N_e \Bigr) + A_{ge} N_e \, .
\ee
For a system in equilibrium, $\delta N_e = \delta N_g = 0$, and so
\be
n_\gamma(\omega) = \frac{A_{ge} N_e}{B_{eg} N_g - B_{ge} N_e} = \frac{\frac{A_{ge}}{B_{ge}}}{\frac{B_{eg}}{B_{ge}} \frac{N_g}{N_e} - 1} \, .
\label{eqdiste}
\ee
Further, in equilibrium the populations of modes with energy $E$ obey Maxwell-Boltzmann distribution, so that 
\be
\frac{N_e}{N_g} = \frac{g_e e^{-E_e/T}}{g_g e^{-E_g/T}} =  \frac{g_e}{g_g} \, e^{- \omega/T} \, . 
\ee
The numbers $g_g$ and $g_e$ are degeneracies (in energy) of the ground and excited states.
Substituting this into (\ref{eqdiste}), we reproduce Planck's formula for photon distribution \underline{iff}
\be
B_{eg} \, g_g = B_{ge} \, g_e \, , ~~~~~~~~~~~~ A_{ge} = \frac{\omega^3}{\pi^2} B_{ge} \, . 
\ee

Thus if we calculate e.g. $B_{ge}$ from first principles, all other parameters are determined. 
We could get a possible leading contribution to $B_{ge}$ by immersing a Hydrogen in a weak, spatially uniform, randomly oriented 
harmonic electric field, with a Planck's distribution of frequencies and a random field orientation. This will excite the atom, inducing 
Rabi oscillations \cite{tong}, with the probability
\be
P_e =  \frac{8\pi \alpha}{3} |\langle e | \vec r | g \rangle|^2 \int d \omega \frac{n_\gamma(\omega)}{(\omega-\omega_0)^2} \sin^2\bigl(\frac{(\omega-\omega_0)t}{2})\bigr) \, ,
\ee
where $E_e - E_g = \omega_0$ and $1/3$ arises from directional averaging. 
After evaluating the integral using contour techniques, the result gives
the width $B_{ge} = dP_e/dt$, 
\be
B_{ge} =  \frac{4\pi^2 \alpha}{3}  |\langle e | \vec r | g \rangle|^2 \, .
\ee
Thus the spontaneous emission rate is
\be
A_{ge} = \frac{4\alpha \omega^3}{3}  |\langle e | \vec r | g \rangle|^2 \, .
\label{spontem}
\ee

Note, that  $|\langle e | \vec r | g \rangle|^2$ could vanish for some states $e$, by symmetry, leading to specific selection rules
allowing or prohibiting some transitions. However this does not imply the stability of the state $e$; instead, other channels
can occur, mediated by higher multipoles in the expansion, or via indirect processes $e \rightarrow e' \rightarrow \ldots \rightarrow g$,
whereby $e$ decays to the ground state in steps. 
The only stable state is the ground state\footnote{In \cite{Bohr:1913zba}, Bohr 
referred to it as the ``permanent" state.}. 

Thus any excited Hydrogen atom, and for that matter all excited Hydrogen atoms in our universe, left to their own facilities, 
decay back to the ground state - 
providing perfect oriented clocks which use the background time as the clock parameter, and give it a direction. The decay selects and keeps the 
arrow of time. This is our main point here:

\vskip.3cm
\noindent {\it Once a system picked the background time, and it has a unique stable ground state, which can only happen 
when there is ${\cal N} \rightarrow \infty$ states in the Hilbert space, the system will also pick and preserve the arrow of time.}
\vskip.3cm

Had the Hamiltonian not been bounded from below, the decay process to ever lower energy states would have 
continued indefinitely, and so the time reversal $t \rightarrow -\infty$ would have 
looked just like the original tower of transitions. However just having a lowest energy state is not sufficient to guarantee time's arrow.
A case in point is precisely the aforementioned Jaynes-Cummings model \cite{Jaynes:1963zz}, which describes a two-state ``atom"
interacting with a quantized ``photon field" of a single fixed frequency (and a discretely infinite spectrum above it), 
enclosed in a cavity. Even though the system has a spectrum bounded
from below, it does {\it not} have a definite arrow of time. The Hamiltonian eigenfunctions of this system are not the atom
``flavor" eigenfunctions, and so during evolution an atom ground state can 
spontaneously interact with the photon and become excited again, leading to the periodic process
of quantum collapse and quantum revival \cite{Eberly:1980zz}. Heuristically, one might think of this as being due to the previously emitted
photons, which never leave the cavity, ``bounce" off the walls, and every so often re-excite an atom in the ground state.

As the cavity walls are moved out the revival delay increases. Eventually, with the full removal of the wall, and restoration of the
full continuum of free photons which can escape to infinity, there will be no revival. 
The Hydrogen ground state will then be a quantum attractor of evolution, and together with a huge number of decay channels taking
energy to infinity, it will manifestly remove any semblance of universal $t \leftrightarrow -t$ symmetry -- finally 
inducing a definite \underline{arrow of time}.

Further, note that in the formula (\ref{spontem}) all the reference to thermal equilibrium and 
the reservoir of photons has completely disappeared. The 
equation is completely microscopic and local, applying to a single isolated Hydrogen atom. Many authors in quantum field theory
have stressed this repeatedly, based on the original papers calculating spontaneous emission rate in quantum field theory
\cite{Dirac:1927dy,Weisskopf:1930au,Weisskopf:1930ps}. This is reflected in the textbooks and lectures, too \cite{lamb,gerryknight,tong}. 
While it is certainly true that (\ref{spontem}) can be correctly calculated without ever invoking statistical notions, 
Einstein's prescient argument  \cite{Einstein:1917zz} feels too elegant to be a mere accident. Since the consistent description 
of Hydrogen requires deploying coherent photon states \cite{Glauber:1963tx,Klauder:1995yr}, it is tempting to imagine that the 
description of the evolution of the coherent state describing the photon sector in spontaneous emission could be
the process of entanglement initiated by the emitted photon evacuating to infinity, in a way proposed in 
\cite{Reimann:2008oeh,Linden:2008awz} to describe the equilibration of a complex quantum system. In these papers, the authors study 
complex quantum systems with many ingredients and show that arrow of time arises dynamically for systems which are in contact
with a large external thermal bath. Here, we have given the argument that even an isolated system, a single atom, behaves in this way, without
an explicitly present external heat bath. Instead, we think the spatial infinity plays the role of the heat sink, of sorts.  
Since the coherent state 
involves contributions from many photons, an excited atom `thermalizes' with these virtual photons, and this process is
irreversible. While microscopic QFT calculations of the decay probability (\ref{spontem}) obscure this point -- manifestly involving only the
electron and a single emitted photon, Einstein's indirect calculation hints that the process of spontaneous emission is indeed akin
to thermalization. Putting this on firmer mathematical grounds seems like an intriguing avenue to explore.

\section{Discussion and Summary}

While we have focused above on Hydrogen, it should be clear that the conclusions apply equally well to other atoms and molecules.
As long as they have Hamiltonians bounded from below and interact with photons (or other continua of mediators), the ground state
will be the quantum attractor. So if we ignore gravity, 
we can imagine an ensemble of such clocks distributed around the universe at any particular time,
as defined by any particular observer, and they will all locally pick the same arrow of time, as dictated by evolution. 

\begin{figure}[thb]
    \centering
    \includegraphics[width=.42\textwidth]{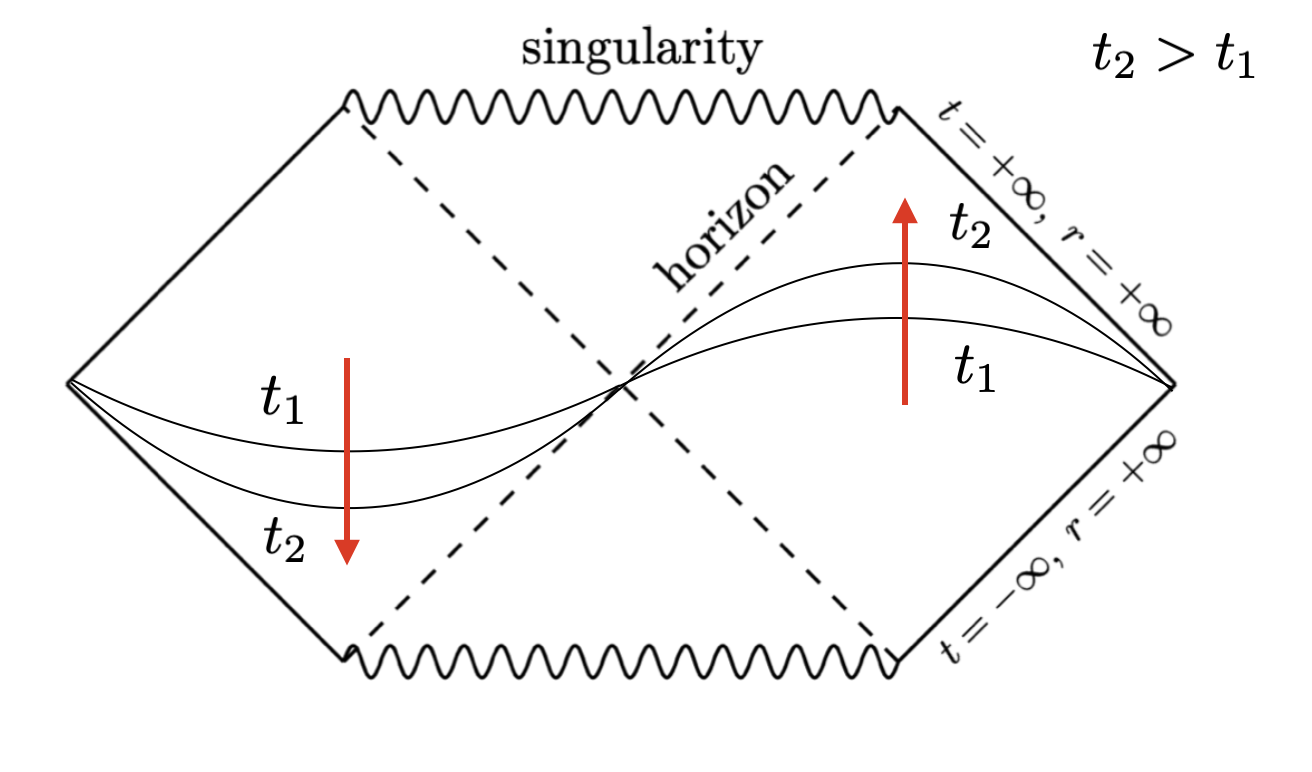}
          \caption{Schwarzschild time reversal, with the local time arrows shown in red.}
    \label{fig2}
\end{figure}
If on the other hand
we include gravity, we can wonder if by making the distribution of clocks very dense somewhere we might get gravitational effects to tilt, and maybe even
invert time's arrow. That can indeed happen; perhaps the simplest example is Schwarzschild black hole,
where the gravitational effects flip the coordinate time direction from one asymptotic infinity to another. However the observers
who are experiencing opposite time arrows are separated by event horizons, and so they cannot causally communicate the flip to each other,
at least classically. Quantum mechanically, the answer is still behind the veils covering Hawking's radiation \cite{Hawking:1974rv}. 
This behavior seems universal under generic conditions, given Hawking's chronology protection conjecture \cite{Hawking:1991nk}.
While gravity can flip time's arrow along spacelike surfaces, classical observers cannot see that due to the causal barriers posed by horizons. 

Time's arrow in cosmology and its connection to inflation is another set of open questions  \cite{Gold:1958koa,Penrose:1979azm,Davies:1983nf,Page:1983uh,Davies:1984qc,Penrose:1988mg,Albrecht:2002uz,Carroll:2004pn}. A big additional 
problem which adds to the confusion is that the analogue to the Hamiltonian with gravity included is the ADM Hamiltonian density
which vanishes identically for {\it all} spacetimes: ${\cal H}_{ADM} = 0$ \cite{Arnowitt:1962hi}. Thus there is no obvious candidate even for 
background time,
let alone its arrow. Yet, we have argued that this problem might not be all that different from the situation with time in ordinary quantum mechanics of 
completely degenerate systems, since in that case the shift of the Hamiltonian can set it to zero on all of the Hilbert space. 

Given this remarkable similarity, as well as the fact that realistic quantum systems with unique stable ground states feature spontaneous emission,
it is conceivable that in the subspace of de Sitter spaces, in theories where cosmological constant can be screened by fluxes of $4$-forms, an arrow
of time also arises due to decay. In this case there is a tendency of a de Sitter space of high curvature to decay to a de Sitter space of lower curvature  \cite{Brown:1987dd,Brown:1988kg}. Under certain 
quite general conditions \cite{Kaloper:2022oqv,Kaloper:2022yiw,Kaloper:2022jpv,Kaloper:2023xfl}, 
when the flux terms screening the cosmological constant are dominated by linear terms,
the vanishing cosmological constant and the associated Minkowski space are the unique quantum attractor. This behavior is analogous to
the ground state of Hydrogen. This feature might well be a factor in the emergence of time's arrow in at least spacetimes which include de Sitter 
sections. In this case, the universe could start as a rare random 
fluctuation \cite{Tryon:1973xi,Garriga:1997ef} from flat space \cite{Carroll:2004pn}, which unlike the spontaneous excitement
of an atom could occur because ${\cal H}_{ADM} = 0$, so that there is no energy barrier, 
but then it evolves back toward Minkowski. This might correlate the smallness of the cosmological constant with cosmic time's arrow.
We think that these are interesting issues, and intend to return to them elsewhere. 
%``Time Is on My Side" comes to mind.

\vskip1cm

{\bf Acknowledgments}: 
We would like to thank A. Albrecht, G. D'Amico, and A. Westphal  
for useful discussions. NK is supported in part by the DOE Grant DE-SC0009999.


\begin{thebibliography}{99}

%\cite{Gold:1958koa}
\bibitem{Gold:1958koa}
T.~Gold,
%``The arrow of time,''
%0 citations counted in INSPIRE as of 18 May 2023
American Journal of Physics 30, 403-410 (1962).

%\cite{Penrose:1979azm}
\bibitem{Penrose:1979azm}
R.~Penrose,
``Singularities and time-asymmetry,'', in {\it General Relativity: An Einstein Centenary Survey},
edited by S. Hawking and W. Israel, Cambridge University Press, Cambridge UK 1979.
%7 citations counted in INSPIRE as of 20 Jul 2022
 
%\cite{Davies:1983nf} 
\bibitem{Davies:1983nf}
P.~C.~W.~Davies,
%``Inflation and Time Asymmetry: Or What Wound Up the Universe?,''
Nature \textbf{301}, 398-400 (1983). 
%doi:10.1038/301398a0
%31 citations counted in INSPIRE as of 23 Jul 2022

%\cite{Page:1983uh}
\bibitem{Page:1983uh}
D.~N.~Page,
%``INFLATION DOES NOT EXPLAIN TIME ASYMMETRY,''
Nature \textbf{304}, 39-41 (1983). 
%doi:10.1038/304039a0
%25 citations counted in INSPIRE as of 20 Jul 2022

%\cite{Davies:1984qc}
\bibitem{Davies:1984qc}
P.~C.~W.~Davies,
%``Inflation and Time Asymmetry,''
Nature \textbf{312}, 524-527 (1984).
%doi:10.1038/312524a0
%7 citations counted in INSPIRE as of 23 Jul 2022
 
%\cite{Penrose:1988mg}
\bibitem{Penrose:1988mg}
R.~Penrose,
%``Difficulties with inflationary cosmology,''
Annals N. Y. Acad. Sci. \textbf{571}, 249-264 (1989).
%84 citations counted in INSPIRE as of 20 Jul 2022

%\cite{Albrecht:2002uz}
\bibitem{Albrecht:2002uz}
A.~Albrecht,
%``Cosmic inflation and the arrow of time,''
[arXiv:astro-ph/0210527 [astro-ph]].
%54 citations counted in INSPIRE as of 20 Jul 2022

%\cite{Carroll:2004pn}
\bibitem{Carroll:2004pn}
S.~M.~Carroll and J.~Chen,
%``Spontaneous inflation and the origin of the arrow of time,''
[arXiv:hep-th/0410270 [hep-th]].
%107 citations counted in INSPIRE as of 01 Feb 2022

\bibitem{Yukalov}
V.~I.~Yukalov,
%``Irreversibility of time for quasi-isolated systems,''
Phys. Lett. A \textbf{308}, 313-318 (2003)
[arXiv:cond-mat/0303497 [cond-mat]]. 

%\cite{Donoghue:2020mdd}
\bibitem{Donoghue:2020mdd}
J.~F.~Donoghue and G.~Menezes,
%``Quantum causality and the arrows of time and thermodynamics,''
Prog. Part. Nucl. Phys. \textbf{115}, 103812 (2020)
%doi:10.1016/j.ppnp.2020.103812
[arXiv:2003.09047 [quant-ph]].
%16 citations counted in INSPIRE as of 30 May 2023

%\cite{Frohlich:2022yir}
\bibitem{Frohlich:2022yir}
J.~Fr\"ohlich,
%``Irreversibility and the Arrow of Time,''
[arXiv:2202.04619 [quant-ph]].
%0 citations counted in INSPIRE as of 30 May 2023

%\cite{Isham:1992ms}
\bibitem{Isham:1992ms}
C.~J.~Isham,
%``Canonical quantum gravity and the problem of time,''
NATO Sci. Ser. C \textbf{409}, 157-287 (1993)
[arXiv:gr-qc/9210011 [gr-qc]].
%540 citations counted in INSPIRE as of 01 Aug 2023

\bibitem{messiah}
A. Messiah, {\it Quantum Mechanics vols. I $\&$ II}, Dover, Mineola, NY 1999. 

%\cite{Arnowitt:1962hi}
\bibitem{Arnowitt:1962hi}
R.~L.~Arnowitt, S.~Deser and C.~W.~Misner,
%``The Dynamics of general relativity,''
Gen. Rel. Grav. \textbf{40}, 1997-2027 (2008)
%doi:10.1007/s10714-008-0661-1
[arXiv:gr-qc/0405109 [gr-qc]].
%1931 citations counted in INSPIRE as of 23 May 2023

%\cite{Susskind:1964zz}
\bibitem{Susskind:1964zz}
L.~Susskind and J.~Glogower,
%``Quantum mechanical phase and time operator,''
Physics Physique Fizika \textbf{1}, no.1, 49-61 (1964). 
%doi:10.1103/PhysicsPhysiqueFizika.1.49
%132 citations counted in INSPIRE as of 18 May 2023

%\cite{Carruthers:1968my}
\bibitem{Carruthers:1968my}
P.~Carruthers and M.~M.~Nieto,
%``Phase and angle variables in quantum mechanics,''
Rev. Mod. Phys. \textbf{40}, 411-440 (1968).
%doi:10.1103/RevModPhys.40.411
%251 citations counted in INSPIRE as of 18 May 2023

%\cite{DiBiagio:2020jbd}
\bibitem{DiBiagio:2020jbd}
A.~Di Biagio, P.~Don\`a and C.~Rovelli,
%``The arrow of time in operational formulations of quantum theory,''
Quantum \textbf{5}, 520 (2021)
%doi:10.22331/q-2021-08-09-520
[arXiv:2010.05734 [quant-ph]].
%17 citations counted in INSPIRE as of 01 Aug 2023

%\cite{Rovelli:2021elq}
\bibitem{Rovelli:2021elq}
C.~Rovelli,
%``The layers that build up the notion of time,''
[arXiv:2105.00540 [gr-qc]].
%2 citations counted in INSPIRE as of 01 Aug 2023

%\cite{Buchholz:2023lhl}
\bibitem{Buchholz:2023lhl}
D.~Buchholz and K.~Fredenhagen,
%``Arrow of time and quantum physics,''
[arXiv:2305.11709 [math-ph]].
%0 citations counted in INSPIRE as of 01 Aug 2023

%\cite{Jaynes:1963zz}
\bibitem{Jaynes:1963zz}
E.~T.~Jaynes and F.~W.~Cummings,
%``Comparison of quantum and semiclassical radiation theories with application to the beam maser,''
IEEE Proc. \textbf{51}, 89-109 (1963). 
%doi:10.1109/PROC.1963.1664. 
%657 citations counted in INSPIRE as of 21 May 2023

%\cite{Eberly:1980zz}
\bibitem{Eberly:1980zz}
J.~H.~Eberly, N.~B.~Narozhny and J.~J.~Sanchez-Mondragon,
%``Periodic Spontaneous Collapse and Revival in a Simple Quantum Model,''
Phys. Rev. Lett. \textbf{44}, 1323-1326 (1980).
%doi:10.1103/PhysRevLett.44.1323
%108 citations counted in INSPIRE as of 21 May 2023

%\cite{Dyson:2002pf}
\bibitem{Dyson:2002pf}
L.~Dyson, M.~Kleban and L.~Susskind,
%``Disturbing implications of a cosmological constant,''
JHEP \textbf{10}, 011 (2002)
%doi:10.1088/1126-6708/2002/10/011
[arXiv:hep-th/0208013 [hep-th]].
%353 citations counted in INSPIRE as of 23 May 2023

%\cite{Glauber:1963tx} 
\bibitem{Glauber:1963tx}
R.~J.~Glauber,
%``Coherent and incoherent states of the radiation field,''
Phys. Rev. \textbf{131}, 2766-2788 (1963). 
%doi:10.1103/PhysRev.131.2766
%1368 citations counted in INSPIRE as of 19 May 2023

%\cite{Klauder:1995yr}
\bibitem{Klauder:1995yr}
J.~R.~Klauder,
%``Coherent states for the hydrogen atom,''
J. Phys. A \textbf{29}, L293-L298 (1996)
%doi:10.1088/0305-4470/29/12/002
[arXiv:quant-ph/9511033 [quant-ph]].
%33 citations counted in INSPIRE as of 19 May 2023

\bibitem{meitner}
L.~Meitner, 
%``†ber die Entstehung der ?-Strahl-Spektren radioaktiver Substanzen."
Z. Physik 9, 131-144 (1922). 
%https://doi.org/10.1007/BF01326962

\bibitem{auger}
P.~Auger,
%``Sur les rayons ? secondaires produits dans un gaz par des rayons X,"
C.R.A.S. 177, 169-171 (1923).

%\cite{Dirac:1927dy}
\bibitem{Dirac:1927dy}
P.~A.~M.~Dirac,
%``Quantum theory of emission and absorption of radiation,''
Proc. Roy. Soc. Lond. A \textbf{114}, 243 (1927).
%doi:10.1098/rspa.1927.0039
%477 citations counted in INSPIRE as of 21 May 2023

%\cite{Weisskopf:1930au}
\bibitem{Weisskopf:1930au}
V.~Weisskopf and E.~P.~Wigner,
%``Calculation of the natural brightness of spectral lines on the basis of Dirac's theory,''
Z. Phys. \textbf{63}, 54-73 (1930). 
%doi:10.1007/BF01336768
%602 citations counted in INSPIRE as of 21 May 2023

%\cite{Weisskopf:1930ps}
\bibitem{Weisskopf:1930ps}
V.~Weisskopf and E.~P.~Wigner,
%``Over the natural line width in the radiation of the harmonius oscillator,''
Z. Phys. \textbf{65}, 18-29 (1930). 
%doi:10.1007/BF01397406
%310 citations counted in INSPIRE as of 22 May 2023

\bibitem{lamb}
M.~Sargent III, M.~O.~Scully, W.~E.~Lamb, {\it Laser Physics},
Addison-Wesley Publishing Co., Reading, MA 1974.

\bibitem{gerryknight}
C.~C.~Gerry, P.~L.~Knight, {\it Introductory Quantum Optics}, 
Cambridge University Press, Cambridge UK 2005. 

%\cite{Einstein:1917zz}
\bibitem{Einstein:1917zz}
A.~Einstein,
%``Zur Quantentheorie der Strahlung,''
Phys. Z. \textbf{18}, 121-128 (1917). 
%153 citations counted in INSPIRE as of 19 May 2023

\bibitem{tong}
D. Tong, {\it Lectures on Applications of Quantum Mechanics}, Part II Cambridge Tripos. 
%, https://www.damtp.cam.ac.uk/user/tong/aqm.html.

%\cite{Bohr:1913zba}
\bibitem{Bohr:1913zba}
N.~Bohr,
%``On the Constitution of Atoms and Molecules,''
Phil. Mag. Ser. 6 \textbf{26}, 1-24 (1913). 
%doi:10.1080/14786441308634955
%126 citations counted in INSPIRE as of 22 May 2023

%\cite{Reimann:2008oeh}
\bibitem{Reimann:2008oeh}
P.~Reimann,
%``Foundation of Statistical Mechanics under Experimentally Realistic Conditions,''
Phys. Rev. Lett. \textbf{101}, 190403 (2008)
%doi:10.1103/PhysRevLett.101.190403
[arXiv:0810.3092 [cond-mat.stat-mech]].
%124 citations counted in INSPIRE as of 23 May 2023

%\cite{Linden:2008awz}
\bibitem{Linden:2008awz}
N.~Linden, S.~Popescu, A.~J.~Short and A.~Winter,
%``Quantum mechanical evolution towards thermal equilibrium,''
Phys. Rev. E \textbf{79}, no.6, 061103 (2009)
%doi:10.1103/PhysRevE.79.061103
[arXiv:0812.2385 [quant-ph]].
%159 citations counted in INSPIRE as of 23 May 2023

%\cite{Hawking:1974rv}
\bibitem{Hawking:1974rv}
S.~W.~Hawking,
%``Black hole explosions,''
Nature \textbf{248}, 30-31 (1974). 
%doi:10.1038/248030a0
%4111 citations counted in INSPIRE as of 23 May 2023

%\cite{Hawking:1991nk}
\bibitem{Hawking:1991nk}
S.~W.~Hawking,
%``The Chronology protection conjecture,''
Phys. Rev. D \textbf{46}, 603-611 (1992). 
%doi:10.1103/PhysRevD.46.603
%597 citations counted in INSPIRE as of 22 May 2023

%\cite{Brown:1987dd}
\bibitem{Brown:1987dd}
J.~D.~Brown and C.~Teitelboim,
%``Dynamical Neutralization of the Cosmological Constant,''
Phys. Lett. B \textbf{195}, 177-182 (1987). 
%doi:10.1016/0370-2693(87)91190-7
%296 citations counted in INSPIRE as of 23 May 2023

%\cite{Brown:1988kg}
\bibitem{Brown:1988kg}
J.~D.~Brown and C.~Teitelboim,
%``Neutralization of the Cosmological Constant by Membrane Creation,''
Nucl. Phys. B \textbf{297}, 787-836 (1988).
%doi:10.1016/0550-3213(88)90559-7
%393 citations counted in INSPIRE as of 23 May 2023

%\cite{Kaloper:2022oqv}
\bibitem{Kaloper:2022oqv}
N.~Kaloper,
%``Hidden variables of gravity and geometry and the cosmological constant problem,''
Phys. Rev. D \textbf{106}, no.6, 065009 (2022)
%doi:10.1103/PhysRevD.106.065009
[arXiv:2202.06977 [hep-th]].
%11 citations counted in INSPIRE as of 19 May 2023

%\cite{Kaloper:2022yiw}
\bibitem{Kaloper:2022yiw}
N.~Kaloper,
%``General relativity on the multiverse and nature\textquoteright{}s hierarchies,''
Phys. Rev. D \textbf{106}, no.4, 044023 (2022)
%doi:10.1103/PhysRevD.106.044023
[arXiv:2202.08860 [hep-th]].
%9 citations counted in INSPIRE as of 19 May 2023

%\cite{Kaloper:2022jpv}
\bibitem{Kaloper:2022jpv}
N.~Kaloper and A.~Westphal,
%``Quantum-mechanical mechanism for reducing the cosmological constant,''
Phys. Rev. D \textbf{106}, no.10, L101701 (2022)
%doi:10.1103/PhysRevD.106.L101701
[arXiv:2204.13124 [hep-th]].
%9 citations counted in INSPIRE as of 19 May 2023

%\cite{Kaloper:2023xfl}
\bibitem{Kaloper:2023xfl}
N.~Kaloper,
%``de Sitter Space Decay and Cosmological Constant Relaxation in Braney Unimodular Gravity,''
[arXiv:2305.02349 [hep-th]].
%0 citations counted in INSPIRE as of 19 May 2023

%\cite{Tryon:1973xi}
\bibitem{Tryon:1973xi}
E.~P.~Tryon,
%``Is the universe a vacuum fluctuation,''
Nature \textbf{246}, 396 (1973). 
%doi:10.1038/246396a0
%229 citations counted in INSPIRE as of 23 May 2023

%\cite{Garriga:1997ef}
\bibitem{Garriga:1997ef}
J.~Garriga and A.~Vilenkin,
%``Recycling universe,''
Phys. Rev. D \textbf{57}, 2230-2244 (1998)
%doi:10.1103/PhysRevD.57.2230
[arXiv:astro-ph/9707292 [astro-ph]].
%152 citations counted in INSPIRE as of 22 Apr 2023

\end{thebibliography}
\end{document}